\documentclass[referee]{aa}

\usepackage{natbib}

\usepackage{graphicx}
\usepackage{amsmath}
\usepackage{amssymb}

 \titlerunning{Morphology and density of post-CME current sheets}
  \authorrunning{Vr\v{s}nak et al.}

\begin{document}
  \title{Morphology and density structure of post-CME current sheets }
%
   \author{ B.~Vr\v{s}nak\inst{1} \and G. Poletto\inst{2}
      \and E. Vuji\'c\inst{3} \and A. Vourlidas\inst{4}
      \and Y.-K. Ko\inst{4} \and J. C. Raymond\inst{5} \and A. Ciaravella\inst{6}
       \and T. \v{Z}ic\inst{1}
      \and D. F. Webb\inst{7} \and A. Bemporad\inst{8}  \and F. Landini\inst{9}
       \and G. Schettino\inst{9} \and C. Jacobs\inst{10}  \and S. T. Suess\inst{11}
           }
  \offprints{B.~Vr\v{s}nak, \email{bvrsnak@geof.hr}}
   \institute{
 Hvar Observatory, Faculty of Geodesy, Zagreb, Croatia  
            \and
  INAF-Arcetri Observatory, Firenze, Italy  
        \and
   Faculty of Science, Geophysical Department, Croatia 
       \and
 Naval Research Laboratory, Washington DC, USA      
             \and
  Harvard-Smithsonian Center for Astrophysics, Cambridge, USA  
           \and
  INAF-Palermo Observatory, Palermo, Italy  
          \and
    Boston College and AFRL, Hanscom, USA   
          \and
    INAF-Torino Astrophysical Observatory, Pino Torinese, Italy
          \and
  Dept. of Astronomy and Space Science, University of Florence, Italy     
    \and
    Centrum voor Plasma-Astrofysica, K. U. Leuven, Belgium   
        \and
      NASA Marshall Space Flight Center, Huntsville, USA   
             }
    \date{Received 22 August 2008 / Accepted 6 February 2009}
 \abstract{}{}{}{}{}
%
  \abstract
     {Eruption of a coronal mass ejection (CME) drags and ``opens" the coronal magnetic field,
presumably leading to the formation of a large-scale current sheet and the field relaxation by
magnetic reconnection.}
     {We analyze physical characteristics of ray-like coronal features
formed in the aftermath of CMEs, to check if the interpretation of this phenomenon in terms of
reconnecting current sheet is consistent with the observations. }
     {The study is focused on measurements of the ray width, density excess, and coronal velocity
field as a function of the radial distance.}
     { The morphology of rays indicates that they occur as a consequence of Petschek-like
reconnection in the large scale current sheet formed in the wake of CME. The hypothesis is
supported by the flow pattern, often showing outflows along the ray, and sometimes also inflows
into the ray. The inferred inflow velocities range from 3 to 30~km\,s$^{-1}$, consistent with the
narrow opening-angle of rays, adding up to a few degrees. The density of rays is an order of
magnitude larger than in the ambient corona. The density-excess measurements are compared with the
results of the analytical model in which the Petschek-like reconnection geometry is applied to the
vertical current sheet, taking into account the decrease of the external coronal density and
magnetic field with height.}
     {The model results are consistent with the observations, revealing that the main cause
of the density excess in rays is a transport of the dense plasma from lower to larger heights by
the reconnection outflow.}


   \maketitle

\section{Introduction}

According to the current comprehension of solar coronal mass ejections (CMEs), the eruption of an
unstable magnetic structure is tightly associated with the formation of the large scale current
sheet (hereinafter CS) in the wake of the eruption. This concept, connecting the large scale
eruption and the localized energy release in a form of flare, was put forward and initially
developed by Carmichael~\citeyear{carmi64}; Sturrock~\citeyear{sturrock66};
Hirayama~\citeyear{hirayama74}; Kopp \& Pneuman~\citeyear{K&P76} (the so-called CSHKP model). When
the current sheet becomes long enough, the tearing instability sets in, leading to fast
reconnection of magnetic field \citep[for the analytical, numerical, laboratory, and observational
results see, e.g.,][ respectively]{furth63,ugai87,gekelman88,vrs03vertical}. Reconnection in the
post-CME current sheet results in an abrupt energy release causing a flare, and on the other hand,
enhances and prolongs the acceleration of the erupting magnetic field structure \citep[e.g.,][ and
references therein]{lin04,vrs08}. Numerical MHD simulations covering various scales, from low
corona up to 1~AU, also demonstrated post-CME current sheet formation, clearly revealing the
importance of the reconnection in the initiation, acceleration, and propagation of CMEs \citep[see,
e.g.,][ for a review see Forbes et al. 2006]{riley02,roussev03,torok04,riley07}.

The most prominent consequence of CME-associated reconnection is the appearance of the so called
two-ribbon flare, closely synchronized with the CME acceleration stage \citep[e.g.,][and references
therein]{maricic07}. The ribbon expansion away from the magnetic inversion line, associated with
growing system of hot loops connecting the ribbons, has led to the formulation of the CSHKP model.
The discovery of cusped structure above the flare loops
\citep[e.g.,][]{tsuneta92,forbes96,tsuneta96}, looptop hard X-ray sources
\citep[e.g.,][]{masuda94,aschwanden96,sui04,veronig06}, the loop shrinkage
\citep[e.g.,][]{svestka87,forbes96,sheeley04,vrs06}, and the recognition of growing posteruption
loop systems in the absence of flares, further supported the CSHKP scenario. Additional evidence
for reconnection below the erupting flux-rope may be found in the so-called disconnection events
\citep[e.g.,][]{webb95,simnett97,wangY99}, downflows above the post-eruption arcades
\citep[e.g.,][]{mckenzie99,innes03,asai04}, horizontal converging flows above the loops
\citep{yokoyama01,lin05}, and flare-associated radio emission below the eruptive prominence
\citep{vrs03vertical}.

Recently, post-CME features seen in UV spectra
\citep{ciara02,ko03,innes03CS,lin05,bemporad06,ciara08}, X-ray images \citep{sui04,sui05}, and
white light coronagraph images \citep[][]{ko03,webb03,lin05} have been attributed to the current
sheets expected to ``connect" the flare loops to the CME core. The UV spectral signature is
generally emission in the high temperature lines of Fe\,{\small XVIII} or Fe\,{\small XIX} formed
at 5\,--\,10 MK. EUV images from the Extreme-ultraviolet Imaging Telescope (EIT) on board the Solar
and Heliospheric Observatory (SoHO) and the Transition Region and Coronal Explorer (TRACE) show
emission in Fe\,{\small XXIV} at even higher temperatures \citep{mckenzie99,innes03}. The UV
features decay slowly, on time scales from several hours \citep{ciara08} up to a few days
\citep[][]{bemporad06,ko03}. The X-ray data from the Reuven Ramaty High Energy Solar Spectroscopic
Imager (RHESSI) show regions of high temperature emission above loop tops, and both the morphology
at different energies and the evolution of the structures supports the CS indentification. The
white light features are bright, narrow rays that appear to map from the cusped flaring loop system
to the CME core \citep[][]{lin05}. Some show blobs of plasma moving outwards at several hundred
km\,s$^{-1}$. Like the UV features, the white light structures can last for a day or two. In
general, white light observations reveal the electron column density, UV data show temperatures,
emission measures and Doppler shifts, and X-ray observations reveal very high temperature plasma.

The aim of this paper is to demonstrate that post-CME rays appear as a consequence of the
reconnection in the current sheet formed in the wake of CME, and to quantify their basic physical
characteristics. We outline the working hypothesis in Sect.~2, whereas details of the model used to
compare the theoretical expectations with observations are provided in the Appendix. Observations
and measurements are presented in Sect.~3, focusing on the morphology and plasma densities in
post-CME rays. The temperature structure and the ionization state will be presented in a separate
paper. Our density measurements are compared with the model estimates and some previous empirical
results in Sect.~4. Results are discussed and conclusions drawn in Sect.~5.

\section{Working hypothesis}

 \begin{figure}
 \includegraphics[width=8.5cm]{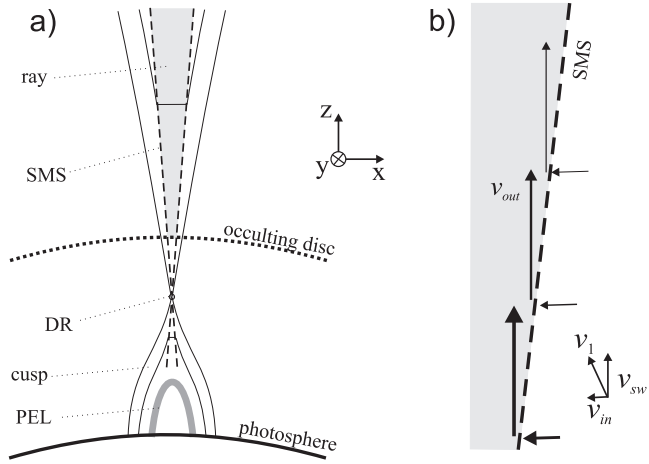}
  \centering
 \caption{
   { a) Interpretation of the ray in terms of the \cite{petschek64} reconnection model:
   DR -- diffusion region; SMS -- slow mode shocks (dashed lines intersecting at DR);
   PEL -- post-eruption loops. Magnetic field-lines are drawn by thin lines; gray area between the
   SMSs outlines the post-CME ray.
   The coordinate system is indicated (the line of sight is in $y$-direction).
 b) An element of bifurcated current sheet; arrows mark plasma flows
($v_{\rm in}$ -- inflow velocity; $v_{\rm sw}$ -- solar wind speed; $v_{\rm out}$ -- outflow
speed), where the arrow thickness depicts plasma densities.
 }
 \label{f1}
 }
\end{figure}

 \begin{figure}
 \includegraphics[width=7cm]{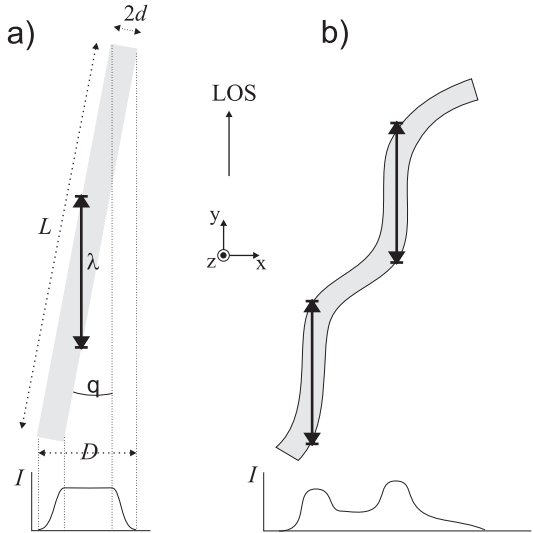}
  \centering
 \caption{
   { Dependence of the width and morphology of the ray at a given height on the orientation/geometry
     of the current sheet: a) straight CS; b) wavy CS. The CS half-thickness is denoted by $d$, horizontal length
     (i.e., the length in $xy$-plane, see also Fig.~1)
     by $L$, the ray width by $D$, and the inclination to the line of sight (LoS) by
     $\theta$. The plasma-column length $\lambda$ is drawn by thick double-arrow. At the bottom
     the intensity profile is sketched.
 }
 \label{f2}
 }
\end{figure}

Since post-CME rays extend outward from cusped structures associated with growing post-eruption
loop-systems, our working hypothesis is that they appear as a consequence of the reconnection in
the CS formed in the wake of CME. Furthermore, we assume the reconnection process can be
approximately described in terms of the steady-state Petschek regime \citep{petschek64}. In this
regime, the reconnection takes place within a small diffusion region (DR), out of which two pairs
of standing slow-mode shocks (SMSs) extend along of the axis of symmetry. Thus, the CS is
bifurcated, with SMSs separating the inflow and outflow region \citep{petschek64,sowardpriest82}.
At SMSs the inflowing plasma is compressed, heated, and accelerated, forming the upward and
downward reconnection jet (Fig.~\ref{f1}). Note that electric currents are concentrated only in DR
and SMSs.

When reconnection takes place in the homogeneous environment, the characteristics of outflowing
plasma are determined primarily by the external plasma-to-magnetic pressure ratio, whereas the
outflow speed is approximately equal to the external Alfv\'en speed \citep[see Appendix
in][]{aurass02fms,skender03}. However, in the case of a vertical current sheet, the ambient coronal
density and magnetic field decrease with height, so the characteristics of the reconnection jets
depend also on the height. The main effect is transport of the dense plasma from lower heights
upward, making the outflow jet much denser than the ambient corona (see Appendix). In this respect,
the post-CME ray should be, to a certain degree, similar to coronal streamers.

Bearing in mind the geometry of the Petschek-type reconnection, the post-CME CS should show
distinct morphological characteristics. It should be thinnest at the height of the diffusion
region, getting wider with increasing height. In the homogeneous plasma the angle between the SMSs
is determined by the inflow Mach number \citep{sowardpriest82,vrssken05}, the half-angle typically
adding up to several degrees \citep{vrssken05}. Thus, for the CS length of several solar radii, the
Petschek-CS thickness (the distance between the SMSs) is on the order of 100 Mm. Note that in the
case of the vertical CS in the solar corona, where the magnetic field diverges radially, the angle
between SMSs should increase with the height even at uniform inflow Mach number (see Appendix), so
the CS thickness should be somewhat larger than in the plane-symmetric case.

Further important characteristics of the post-CME CS that determine how it should look in
coronagraphic images, are the horizontal length and the orientation with respect to the line of
sight (Fig.~\ref{f2}). In this respect, it is important to note that the angle $\theta$ between the
plane of CS and the line of sight (LoS) has to be rather small, otherwise the CS would not appear
as a distinct feature in coronagraphic images. This causes a ``selection effect", which can explain
the fact that post-CME rays can be recognized only in a relatively small fraction of events
\citep{webb03}.

In Fig.~2a we depict how the effective thickness $D$ of the ray depends on the CS thickness $2d$,
horizontal length $L$, and the angle $\theta$. On average, a ray should be thinner and brighter for
smaller $\theta$ (larger LoS column-length $\lambda$). In Fig.~\ref{f2}b we illustrate how the
variation of $\theta$ along the LoS can lead to a complex appearance of the ray \citep[see
also][]{saez07}, i.e., a wavy CS would be seen by observer as a ``multi-ray" structure. Note that
the angle $\theta$, or its variation along the LoS, can vary in time and height, so the morphology
of the structure can change in time and can be different at different heights.

\section{Observations}

\subsection{SoHO/LASCO mass images}

For the analysis of the morphology and the column-density of post-CME rays we employ data provided
by the Large Angle Spectroscopic Coronagraph \citep[LASCO;][]{brueckner95} on board SoHO. We focus
on the LASCO-C2 images, covering the radial distance range from 2 to 6 solar radii ($r_{\odot}$).
In particular, we use the so-called mass images that are derived from LASCO base-difference images
calibrated in units of solar brightness. These images show changes of the column-mass along the
line of sight under the assumption that the structure lies in the plane of sky \citep[for details
we refer to][]{billings66,poland81,vourlidas00,vourlidas02}; bright pixels (positive values)
represent areas where the column-mass is increased, whereas dark pixels (negative values) show
depleted regions.

The initial sample of post-CME rays considered for the analysis consisted of eleven events. We
emphasize that these rays were spotted while analyzing the associated CMEs for other purposes,
i.e., the rays were chosen rather randomly. Thus, no attempt was made to analyze the occurrence
rate of post-CME rays \citep[for the criteria and statistical background we refer to][]{webb03}.
Nevertheless, it should be noted that ray-like features in the wake of CMEs are not uncommon,
particularly bearing in mind specific circumstances under which a post-CME current sheet could be
observed. That is, the level of post-CME activity should not be too high, current sheet has to be
oriented at small angle with respect to the line of sight, it should be relatively stable, etc.

Inspecting the selected sample we noticed intermittent activity in most of the rays, manifested as
changes of their shape, contrast, inclination, etc. These changes were either caused by coronal
disturbances coming from remote eruptions, or were revealing internal activity, such as blob
formation/ejection and/or outward propagating wave-like perturbations.

For a detailed analysis we chose three post-CME rays, observed on 8-9 January 2002, 18 November
2003, and 26 June 2005. The events of 18 November 2003 and 26 June 2005 were selected as examples
of relatively stable rays. In both events we analyzed only one ray image (12:50~UT and 06:30~UT,
respectively). On the other hand, the event of 8--9 January 2002 was showing a significant
activity, similar to that in the remaining eight events. This event was selected to quantify the
level of variability in post-CME rays, so we performed measurements at three different times
(00:06, 06:06, and 12:06~UT). Note that rays in the selected exposures did not show significant
blob-like features or similar inhomogeneities, although they might have been present in earlier
and/or later times.

\subsection{The ray morphology}

 \begin{figure}
 \includegraphics[width=8.5cm]{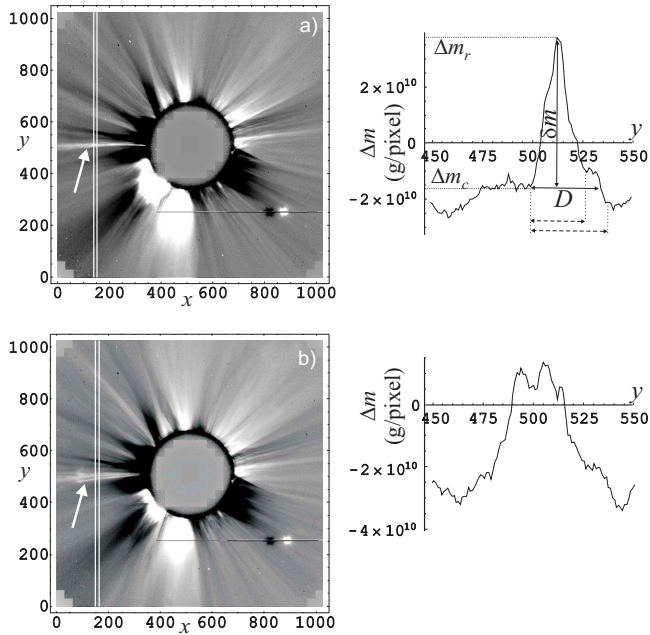}
  \centering
 \caption{
   {  a) Simple post-CME ray (9 January 2002 at 06:06~UT);
    b) Complex ray consisting of three radial substructures (9 January 2002 at 02:30~UT).
   The ray is marked by white arrow.
   The scale of the LASCO-C2 images
   is given in pixels (1 pixel corresponds to 11.9 arcsec).
   The column-mass profiles measured along the bin marked by white lines
   are shown in the right-hand panels (abscissa scale in pixels).
   Estimates of the ray width, $D$, and column-mass excess, $\delta m$, are indicated in a).
   Dashed arrows indicate the lower and upper limit of $D$.
 }
 \label{f3}
 }
\end{figure}

 \begin{figure*}
 \includegraphics[width=18cm]{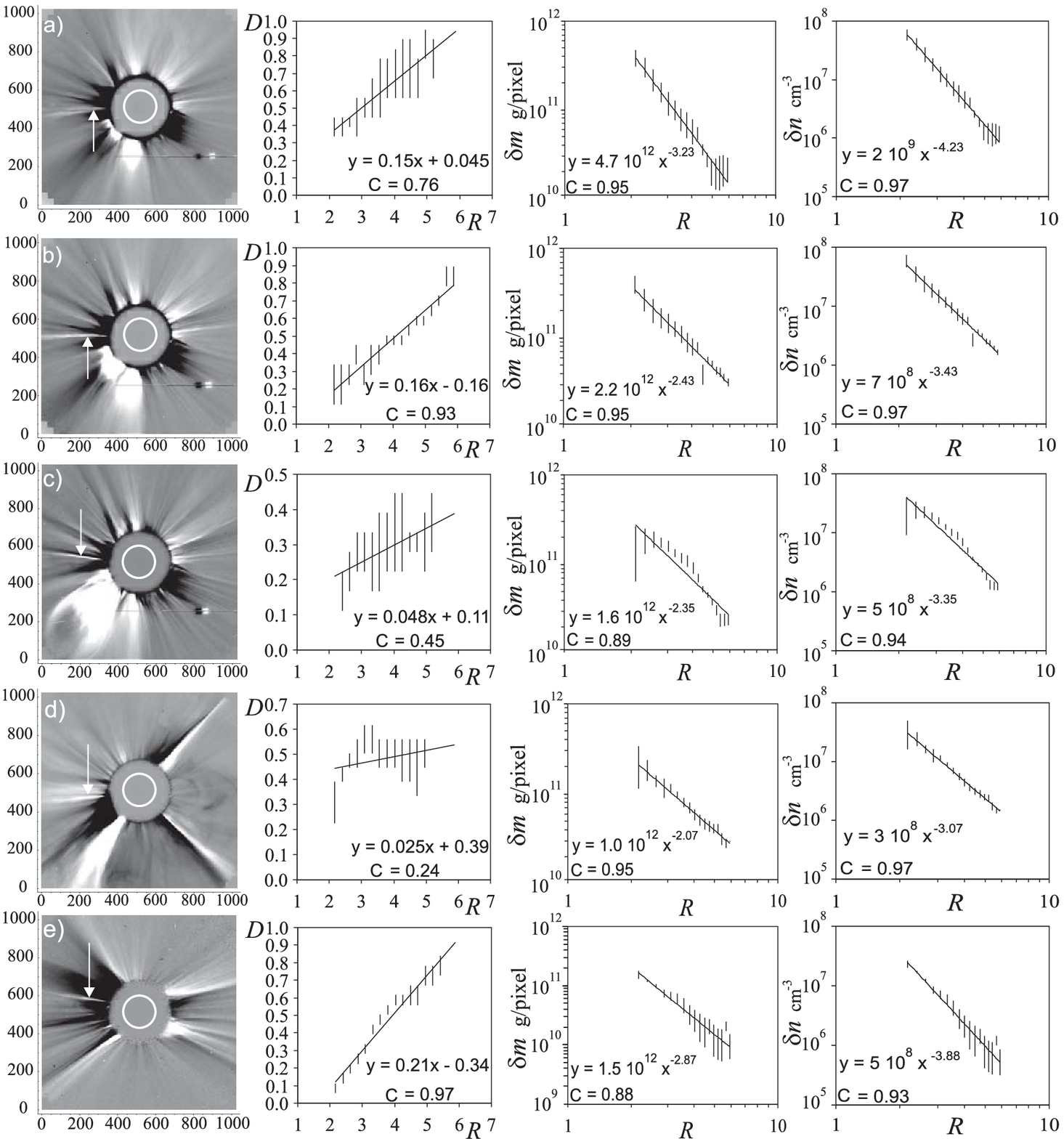}
  \centering
 \caption{
   {  The five measured post-CME rays (indicated by arrows in the full-resolution LASCO-C2 images shown
   in the first column):
   a)\,--\,c) 9 January 2002 at 00:06, 06:06, and 12:06~UT, respectively; d) 18 November 2003 at 12:50~UT; e) 26 June 2005 at
   06:30~UT.
   In the second, third, and fourth column we show the ray width, column-mass excess, and the density excess,
   as a function of the radial distance. The widths and radial distances are expressed in units of the solar radius.
   The least-square fits are shown in the insets, together with the correlation coefficient $C$.
 }
 \label{f4}
 }
\end{figure*}

In Fig.~\ref{f3} we illustrate the variability of the ray morphology by showing two images of the
ray that became recognizable a few hours after the east-limb CME eruption on 8 January 2002 (first
appearance in LASCO-C2 at 17:54 UT). In Fig.~\ref{f3}a we first show a simple ray pattern that was
recorded half a day after the eruption. Four hours earlier, the ray had a complex form consisting
of three radial substructures (Fig.~\ref{f3}b); this ``multi-ray" pattern could be a result of a
wavy current sheet, as illustrated in Fig.~\ref{f2}b. Morphological changes of this kind,
developing on the time scale of hours are a common characteristic of post-CME rays, since they were
present in nine out of eleven events from our initial sample.

The estimates of the ray width and the column-mass excess are based on the LASCO-C2 mass-images
(Fig.~\ref{f3} left), which provide for each pixel the difference $\Delta m$ between the
column-mass in the actual image and the reference image. In the right-hand panels of Fig.~\ref{f3}
we show two mass-difference profiles $\Delta m(y)$, where $\Delta m$ is expressed in g/pixel, while
the abscissa, corresponding to the $y$ coordinate of LASCO images, is represented in pixels. The
profiles shown were measured in the LASCO-C2 difference-images along the bin marked by white lines
in the left panel of Fig.~\ref{f3}, i.e., roughly perpendicular to the ray direction. The bin is 10
pixels wide, and the profile shows $\Delta m$ averaged over the bin width. Pixels with $\Delta m=0$
represent areas where the column-mass remained the same as in the reference image. Note that the
mass astride the rays is depleted ($\Delta m < 0$), probably due to the CME-associated coronal
expansion, usually seen as coronal dimming when the eruption is launched from regions on the solar
disc. The depletion might be partly caused also by the presumed inflow into the current sheet.

Given the working hypothesis, where the ray is considered to be the reconnection outflow jet of a
large-scale Petschek-like bifurcated current sheet, we are interested in the column-mass excess of
the ray, $\delta m$, defined as the difference between the peak in the ray column-mass $\Delta m_r$
and the column-mass in the adjacent corona $\Delta m_c$. In other words, $\delta m$ determines the
difference of column-mass associated with the current sheet and the inflow region ($\delta m=\Delta
m_r-\Delta m_c$; see the right-hand panel of Fig.~\ref{f3}a). The ray width $D$ is measured as the
full width of the ray structure in the mass-difference profile (Fig.~\ref{f3}a right). The main
source of error in estimating $\delta m$ and $D$ is in determining the value $\Delta m_c$, since
this value is frequently different on opposite sides of the ray (Fig.~\ref{f3}a right). Thus, we
estimated the upper and lower limit (dashed arrows in Fig.~\ref{f3}a right), and these measurements
gave us the error bars plotted in Fig.~\ref{f4}.

The measurements illustrated in Fig.~\ref{f3} were performed in five images of post-CME rays over
the LASCO-C2 range of heights to estimate radial dependence of relevant ray parameters. The
measured rays are marked in the mass-images shown in the left column of Fig.~\ref{f4} (hereinafter
denoted as rays a\,--\,e). In the second column of Fig.~\ref{f4} the ray widths, expressed in units
of the solar radius $r_{\odot}$, are presented as a function of the radial distance, $D(R)$, where
$R=r/r_{\odot}$. The ray in Fig.~\ref{f4}a is characterized by ``multi-ray" structure, so we
measured only the central feature, which was the most prominent element of the structure.

Inspecting the graphs one finds that in the distance range $R=2$\,--\,2.5 the widths span between
0.1 and 0.4~$r_{\odot}$ (mean 0.26). In the range $R=5$\,--\,6 the widths increase to
0.4\,--\,0.9~$r_{\odot}$ (mean 0.7). The widths we found at $R\sim2$ are more than twice larger
than found by \citet[][]{webb03} who analyzed the coronagraph data on board the Solar Maximum
Mission (SMM): converting their values of the width expressed in degrees, into units of the solar
radius at the corresponding heights, we find that the mean width reported by \citet[][see Tabs. 2
and 3 therein]{webb03} is 0.09~$r_{\odot}$ in this height range. The difference may be related to
the lower sensitivity of SMM.

\subsection{The ray density}

In the third column of Fig.~\ref{f4} we show the radial dependencies of column-mass excesses
$\delta m(R)$, expressed in g/pixel. Inspecting the graphs $\delta m(R)$ we find that the slopes of
the fitted power-law functions range from $-2.1$ to $-3.2$. The average value adds up to
$-2.6\pm0.5$.

The column-mass excess $\delta m$ was converted to the number-density excess $\delta n$ by assuming
that, due to the spherical geometry, the column length increases proportionally with the radial
distance, $\lambda\propto R$. In particular, we used $\lambda_0=100$~Mm at $R=R_0=2.16$ (the lowest
height of measurements). The outcome is presented in log-log graphs in the fourth column of
Fig.~\ref{f4}, together with the power law fits. Note that some other choice of $\lambda_0$ would
shift vertically the $\delta n(R)$ dependence, without changing the slope in log-log graph. The
slope would change only if some other radial dependence $\lambda(R)$ would be applied, e.g., the
angle $\theta$ might be height-dependent.

Inspecting the graphs $\delta n(R)$ in Fig.~\ref{f4} we find that the slopes of the fitted
power-law functions range from $-3.1$ to $-4.2$, the average value adding up $-3.6\pm0.5$. [We pay
special attention to the slope of the fit, since it does not depend on the presumed value of
$\lambda_0$, so it can be directly compared with the model results (Sect.~4.1).] In this respect it
is important to note that in the case of the smallest slope in the sample, the ray width becomes
approximately constant beyond $R\approx 3$, indicating a possible decrease of $\theta$ with the
height, i.e., the increase of $\lambda$. This would imply that in this case the densities are
overestimated at larger heights, since $\delta n\propto \delta m/\lambda$, i.e., the true slope of
$\delta n(R)$ is probably somewhat steeper.

 \begin{figure}
 \includegraphics[width=8.5cm]{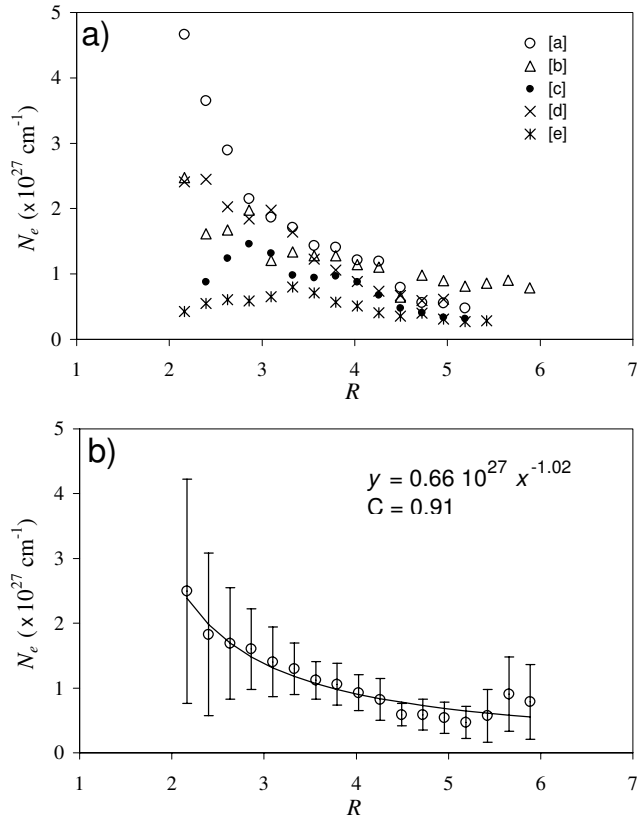}
  \centering
 \caption{
   { Number of electrons per unit length of the ray, $N_e(R)$:
   a) presented separately for the five rays shown in Fig.~\ref{f4}
   (measurements are labeled in the legend, following the sequence in Fig.~\ref{f4});
   b) averaged over the five rays (error bars represent standard deviations), shown together with
   the power-law fit.
 }
 \label{f5}
 }
\end{figure}

Since the evaluation of the density excess depends on the choice of the column length $\lambda$, in
Fig.~\ref{f5} we present an estimate of the number of electrons per unit length of the ray, which
is a parameter independent of $\lambda$. If we approximate the mass-difference profiles across the
ray by triangular profiles, the mass-excess per unit length of the profile amounts to $M=D\delta
m/2$, from which we evaluate the electron-number excess per unit length of the ray, $N_e$.

Inspecting Fig.~\ref{f5}a, where $N_e(R)$ is presented separately for all five rays from
Fig.~\ref{f4}, we find that $N_e(R)$ either decreases (rays [a], [b], and [d]) or stays roughly
constant (rays [c] and [e]). When averaged, the data from Fig.~\ref{f5}b show a dependence similar
to that obtained by \citet{ciara08}, who found that after a decrease at low heights, the value of
$N_e$ becomes approximately constant. Comparing their Fig.~8 with our Fig.~\ref{f5}b we find that
the numbers and the trends are similar. However, the difference is the radial distance at which
$N_e(R)$ becomes constant: according to results by \citet{ciara08} this happens at $R\approx 2$,
whereas in our case the transition occurs at $R\approx 4$. In Fig.~\ref{f5}b we present also the
power-law fit, showing that the overall trend could be described as well by $N_e\propto R^{-1}$.

\subsection{Coronal flows}

 \begin{figure}
 \includegraphics[width=8.5cm]{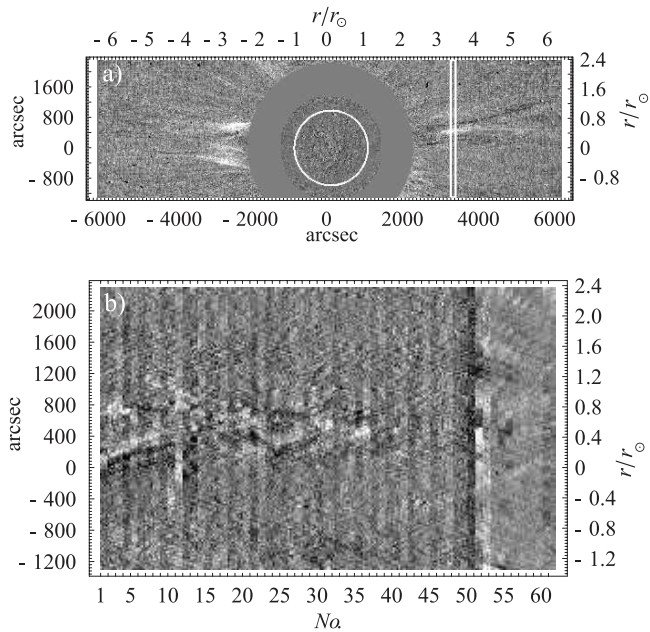}
  \centering
 \caption{
   { a) LASCO-C2 running-difference images of the west-limb ray of 8 January 2002 (13:54~UT), revealing the inflow into the
   ray. The slice used in the stack-plot shown in b) is outlined by white rectangle centered at $x=+3.34\,r_{\odot}$.
   b) Stack-plot composed of slices taken from successive running-difference images; $x$-axis represents the ordinal number
   of the LASCO image for 8 January 2002.
 }
 \label{f6}
 }
\end{figure}

Coronal regions beneath CMEs are generally characterized by highly dynamical intricate structures,
revealing complex magnetoplasma flows and waves. Consequently, it is difficult to find persistent
systematic flow patterns, especially in periods of increased solar activity, when disturbances from
other CMEs affect the coronal region of interest. This also holds for flow patterns associated with
post-CME rays, which show permanent morphological changes and are strongly affected by disturbances
from distant CMEs.

Yet, in certain situations some characteristic flows could be identified. Most frequently,
outward-moving inhomogeneities along the ray are observed, usually having velocities of several
hundred km\,s$^{-1}$ \citep[e.g.,][]{ko03,lin05}. Such motions are often interpreted in terms of
the reconnection outflow, which may or may not be characterized by the Alfv\'enic speed \citep[for
a discussion we refer to][and references therein]{barta07}. In this respect we note that sometimes
signatures of shrinking loops are also observed in the wake of CMEs \citep[e.g.,][and references
therein]{sheeley02,sheeley04,sheeley07}, consistent with decelerated reconnection downflows.

Compared to reconnection outflows, detecting signatures of reconnection inflows is much more
difficult. A possible example was reported by \citet{yokoyama01}, based on the observations by the
EIT/SoHO. Another example was presented by \citet{lin05}, who employed data from EIT and
UltraViolet Coronagraph Spectrometer (UVCS) on board SoHO. Recently, \citet{bemporad08} reported a
``side-reconnection" in the aftermath of a CME, induced by the CME expansion: from the detection of
two converging reconnecting features at 1.7 $r_{\odot}$ (probably the CME flank and the streamer
boundary) the authors inferred an inflow speed of 3\,--\,4 km/s, close to the 5 km/s derived by
\citet{yokoyama01}.

In Fig.~\ref{f6} we present a stack-plot showing an example of the reconnection-inflow pattern
observed above the west limb on 8 January 2002 in the LASCO-C2 field-of-view. Note that this
post-CME feature is not the ray of 8-9 January 2002 presented in Figs.~\ref{f3} and \ref{f4}, which
was on the opposite side of the solar disc. This west-limb event was not observed by UVCS, i.e., we
do not know if it was characterized by the Fe\,{\scriptsize XVIII} emission, which is one of
criteria for identifying post-CME current sheets. It should be stressed that we could not find a
clear/measurable signature of reconnection inflow in the three post-CME rays analyzed in
Sects.~3.2. and 3.3. In these events no suitable inhomogeneities could be identified to provide
tracing of plasma motion in the inflow region (to be discussed in Sect.~5).

 The west-limb ray presented in Fig.~\ref{f6}, formed
after a faint ejection first observed in LASCO-C2 at 02:54 UT on 8 January 2002 (not listed in the
LASCO CME catalog). Each stack represents the cut along a line parallel to the $y$-axis of a given
LASCO-C2 512$\times$512-pixel running-difference image, showing the pixel-intensities in the range
$x=3.30$\,--\,3.38\,$r_{\odot}$. Note that the ray was oriented very close to the $x$-axis of LASCO
images, i.e., the stacks are roughly perpendicular to the ray direction. The stack-plot contains
all images recorded on 8 January 2002 (denoted by ordinal number at the $x-$axis of the stack-plot;
the time cadence is on average 24 min).

The black/white stripes converging towards the ray location, located at $y\sim 400^{\prime\prime}
\sim 0.41 r_{\odot}$, clearly show inflows into the ray. The corresponding velocities in the left
part of the stack-plot are found in the range 15\,--\,25~km\,s$^{-1}$ (mean $19\pm4$~km\,s$^{-1}$).
Such velocities are several times higher than 5~km\,s$^{-1}$ reported by Yokoyama et
al.~(\citeyear{yokoyama01}). After the slice No. 50 (18:54~UT), the stack-plot reveals effects of a
large-scale perturbation from another CME, which occurred above the E-limb. Note that the ``push"
caused by the perturbation increased the inflow speed, now ranging from 25 to 30~km\,s$^{-1}$ (mean
$27\pm5$~km\,s$^{-1}$).

Finally, it should be noted that we also observed intermittent downflows in the form of shrinking
loops \citep[similar to that reported by][]{sheeley04}, starting from $R\sim 3$. At larger heights,
outflows could be recognized, sometimes showing a pattern like in disconnection events
\citep[e.g.,][]{webb95,simnett97,wangY99}. This implies that the diffusion region in this event was
located around $R\sim 3$, i.e., much higher than in the events shown in Fig.~\ref{f4}.

\section{Comparison with the model and previous studies}

\subsection{Comparison with the model results}

 \begin{figure*}
 \includegraphics[width=12cm]{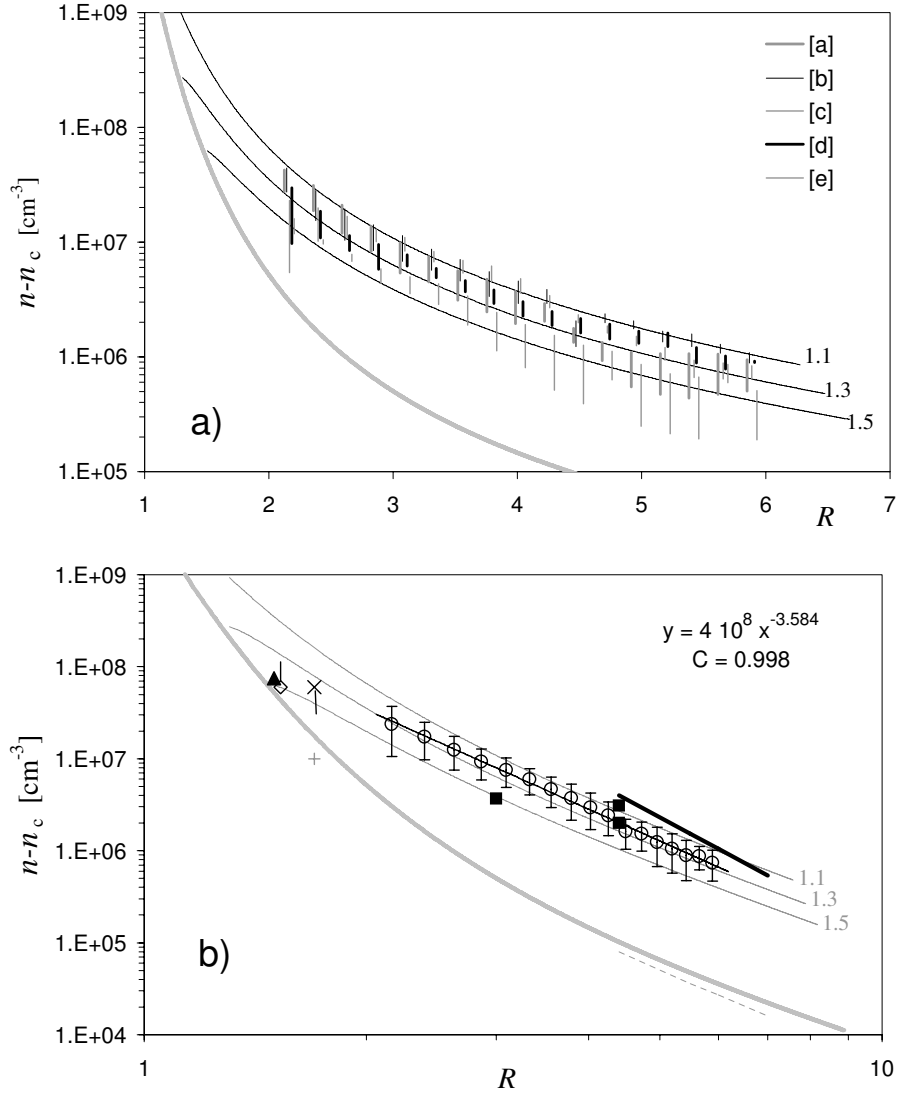}
  \centering
 \caption{
   {  a) Comparison of the density-excess measurements presented in the right hand column of Fig.~\ref{f4} with the
   model results (thin black lines labeled by the heliocentric distance of the diffusion
         region, $R_{\rm x}$). Measurements are labeled in the legend, following the sequence in Fig.~\ref{f4}.
   The thick-gray line shows the density model of a quiet $10^6$~K corona \citep{mann99}.
   b) The mean values of the data presented in (a), shown by circles with error bars and the corresponding power-law fit
   (written in the inset), compared with
   the model results and the values from previous studies reviewed in Sect.~4.2
   (diamond -- \cite{ko03}; triangle -- \cite{ciara02}; squares --  \cite{ko03}; cross -- \citet{bemporad06}).
 }
 \label{f7}
 }
\end{figure*}

In Fig.~\ref{f7}a we present a comparison of measurements presented in 4th column of Fig.~\ref{f4}
with the model-dependencies $\delta n(R)$ derived in the Appendix. In particular, we present the
model results based on the isothermal ($T=10^6$~K) \citet{parker58} solar wind model \citep[see
also][]{mann99}. For the magnetic field $B(R)$ we take the empirical coronal magnetic field scaling
established by \cite{D&M78}. Note that data [a], [b], [d], and [e] are slightly shifted in
$R$-coordinate (symmetrically with respect to [c]) to avoid overlapping. Figure~\ref{f7}a clearly
shows that the observations fit well to the model curves (as stated in Sect.~3.3 the choice of the
column length $\lambda_0$ only shifts the values up or down, but does not change the slope).
Comparing the trend of the data we see that the slope corresponds much better to the model slopes
for the CSs than for the ambient corona. Furthermore one finds that the CS densities are more than
one order of magnitude larger than that in the ambient corona.

In Fig.~\ref{f7}b we present the measured average density excess $\overline {\delta n}(R)$ (circles
with error bars represent the mean values from Fig.~\ref{f7}a with the associated standard
deviations) and the corresponding power-law fit, compared with the CS model dependencies. The
observational data show the dependence $R^{-3.6}$, whereas in the same height range the CS model
data show $\sim$\,$R^{-3.0}$\,--\,$R^{-3.2}$ and the quiet-corona model behaves as $R^{-4.3}$ (see
Appendix).

\subsection{Ray densities from previous studies}

Post-CME rays were studied in a number of papers, employing various techniques to estimate their
density. In Table~\ref{tab1} we present an overview of the spectrographic results obtained from
UltraViolet (UV) spectra of five post-CME rays (dates are given in the header row) reported by
\citet{ciara02}, \citet{ko03}, \citet{raymond03}, \citet{lee06}, \citet{bemporad06}, and
\citet{ciara08}. In the last column we also present the unpublished results by Schettino et al.
(2008, to be submitted).

All rays listed in Table~\ref{tab1} were characterized by the presence of the narrow Fe\,{\small
XVIII} emission. UV spectra of rays have been obtained with the UltraViolet Coronagraph
Spectrometer \citep[UVCS;][]{kohl95} on board SoHO. The spectrometer observes through a narrow
entrance slit 42$^\prime$ long and up to 84$^{\prime\prime}$ wide. The slit can be located at any
polar angle and at heliocentric distances from 1.5 up to 10\,$r_{\odot}$. UVCS can image spectra of
the solar corona in the range 945\,--\,1270 \AA\/ (473\,--\,635 \AA\/ in the second order).

The first five rows of Table 1 specify the associated CME/flare events: the first appearance of the
CME in the LASCO-C2, the CME mean speed, the soft X-ray importance of the flare if observed, the
active region label, and duration of the LASCO-ray if observed, respectively.

The next eight rows concern the ray characteristics as detected in the UV spectra: the time of the
first detection of the ray by UVCS, the height at which the spectrograph slit intersects the ray,
 %
 %
the ray width along the UVCS slit expressed in units of $r_{\odot}$ (both the full-width at
half-maximum and the full width are presented), the estimate of the LoS depth, duration of the ray
UVCS observations, and finally, in last two rows we present the ray temperature and density.

\begin{table*}
 \caption{Spectrographic results on the UV current sheets.   }
\begin{center}
\begin{tabular}{lcccccc}
\hline\\
             & 1998 Mar 23$^1$ & 2002 Jan 8$^2$ & 2002 Apr 21$^3$ & 2002 Nov 26$^4$ & 2003 Nov 4$^5$ & 2003 June 2$^6$ \\
\hline\\
  CME 1st-LASCO                & 09:33:36 & 17:54:05  & 01:27:20 & 17:06:21 & 19:54:05 & 08:54:05 \\
  CME speed (km\,s$^{-1}$)      &   403    &   1794    &   2393   &    479   & 2657  & 980    \\
  Flare                        &   N      &   Ybl$^*$  &   X\,1.5   &    N    & X\,30.8  & M\,3.9   \\
  AR No.                       &    ---     &   9782/85 &   9906   &    ---     & 10486 & 10365    \\
  LASCO CS                     &    N     &   2\,days  &    N    &    N     &  $\sim 20$\,h & 6.5\,h  \\
\hline\\
  CS obs. start                & 16:00:30 & 17:48/Jan\,10  & 00:45:34 & 18:39:15 & 20:03:50 & 09:40:44 \\
  CS obs. height             &   1.51\,(--\,1.38) &  1.53  &   1.62   &   1.61   & 1.69     &1.68   \\
 CS FWHM width ($r_{\odot}$)   &   0.1    &   0.2     &   0.3    &   0.4    & 0.2\,--\,0.1 &  0.3\\
 CS full width ($r_{\odot}$)   &   0.4    &   0.4     &   0.6    &   0.8    & 0.5\,--\,0.2  &  0.35\\
  CS LoS depth ($r_{\odot}$)   &   0.06   &   0.2     &   ---    &   0.5    & 0.1  & 0.1\\
  CS obs. duration             &   20\,h    &   10\,h     &  14\,min  &  2.3\,days & $\sim 17$\,h & 6.3\,h \\
  CS temp. ($10^6$ K)          &    5     &   3\,--\,4  &     5   &   8\,--\,3    & 8\,--\,4 & 7   \\
 CS density($\rm 10^7 cm^{-3}$)& 5\,--\,10  &  $\le$ 4  &   ---  &  6.5\,--\,7.5 & 7\,--\,10 & 1  \\
 \hline
\end{tabular}
\end{center}
\label{tab1}
 1 - \citet{ciara02};\\
  2 - \citet{ko03};\\
   3 - \citet{raymond03}; \citet{lee06};\\
 4 - \citet{bemporad06};\\
  5 - \citet{ciara08}\\
  6 - Schettino et al. (2008, to be submitted)\\
  $^*$ Ybl = Yes (behind the limb)
\end{table*}

The ray UVCS densities are presented in Fig.~\ref{f7}b, together with the density-excess
measurements presented in Sects.~3.3 and 4.1. The estimate based on the UVCS spectra reported by
\cite{ko03} is indicated by the diamond symbol. The column-length was assumed therein to be 140 Mm
which was taken to be the same as the FWHM of the spatial profile of the Fe{\small XVIII}
$\lambda$974 emission across the post-CME rays. Such a column-length is about 2 times larger than
the value we would get for this height by the scaling that was applied in Sect.~3. Thus, to adjust
the values to our $\lambda_0$ the density has to be multiplied by a factor of 2 (indicated by
vertical bar attached to the diamond symbol in Fig.~\ref{f7}b).

The density reported by \cite{ciara02}, also estimated from the UVCS data, is shown by the black
triangle. \cite{ciara02} estimated the CS density at $R=1.5$ to 5\,--\,$10\times10^7$~cm$^{-3}$
assuming the column length of 40~Mm. After adjusting to the column-length equivalent to that we
used in Sect.~3 (required increase is around 20\,\%), we find the equivalent density around
$6\times10^7$~cm$^{-3}$.

The density inferred by \citet{bemporad06} from the spectral observations of the face-on current
sheet of 26 November 2002 is shown by the black cross. The attached vertical bar indicates that the
density could be lower if a larger current sheet thickness would be assumed. The gray plus symbol
represents the estimated density of the ambient coronal plasma.

In Fig.~\ref{f7}b we also show (black squares) the density excess data based on the ray and coronal
densities estimated by \cite{ko03} employing the LASCO-C2 observations and using the Thompson
scattering function \citep{billings66}. In estimating the ray density at $R=4.4$ \cite{ko03}
assumed the column-length of $\sim250$~Mm, which is about 1.2 times larger value than we used at
the same height. Thus, adjusting their results to our column-length would result in 1.2 times lower
density.

Finally, in Fig.~\ref{f7}b the results based on the polarization-brightness measurements performed
by \citet{poletto08} are presented. The thick black line represents the density excess $\delta n$
re-evaluated from Fig.~9 of \citet{poletto08} by applying column-lengths consistent with those
applied in Sect.~3. The ray data drawn by the thick line fit very closely to our measurements,
whereas the ``quiet corona'' data (thin gray dashed line) are very close to the model densities.

The power-law fit to all data (adjusted to the same $\lambda_0$) shown in Fig.~\ref{f7}b has the
slope $R^{-3.3}$ (without the $\lambda_0$ adjustment would add up to $R^{-3.0}$). This is quite
close to the model slopes found in the radial distance range 1.5\,--\,7~$r_{\odot}$, characterized
by the power-law exponents 3.0\,--\,3.2 for $R_{\rm x}=1.1$\,--\,1.5 and the \cite{D&M78} magnetic
field.

\section{Discussion and conclusion}

The empirical characteristics of post-CME rays we summarize as follows:
\begin{enumerate}
\item rays often show activity in the form of outflowing features, morphological changes, and
   changes of the inclination; sometimes reconnection inflows are observed;
\item the width of rays increases with height, from $\sim$\,0.1\,--\,0.3~$r_{\odot}$ at $R\sim 2$ to
   $\sim$\,0.4\,--\,0.8~$r_{\odot}$ at $R\sim 6$.
\item densities found in rays are at least several times (up to more than one order of magnitude)
larger than in the ambient corona in the considered height range;
\item coronal regions surrounding rays are depleted;
\item on average, the number of electrons per unit length of the ray, $N_e$, first decreases with the height,
and then, at larger heights becomes approximately constant;
\item temperature of rays in the range $R\sim 1.5$\,--\,1.7 spans from 3 to 8 MK, i.e.,
it is several times larger than in the ``normal" corona, with a tendency to decrease in time.
\end{enumerate}

The nature of the coronagraphic white-light ray activity is twofold. The formation of fine
structure elements, mostly blob-like features, and their outward motion at speeds in the range
100\,--\,1000~km\,s$^{-1}$, seems to be the most common internal activity. It could be interpreted
in terms of the current sheet tearing, resulting in the formation and outward ejection of plasmoids
\citep[e.g.,][and references therein]{barta07,riley07}. On the other hand, changes of the ray
inclination, as well as morphological changes that could be attributed to changes of the ray
geometry, seem to be caused by the large-scale magnetic field evolution in the wake of the CME. In
some cases this type of activity is caused by perturbations coming from distant eruptions, most
likely the large-amplitude waves or shocks.

As stated in item 2, the width increases on average by about 0.4\,$r_{\odot}$ over the distance of
4\,$r_{\odot}$, i.e., the width is roughly proportional to the radial distance. Following our
interpretation of rays in terms of the reconnecting current sheet, the ray boundaries should
outline the slow mode shocks. Given the model presented in the Appendix, in particular
Fig.~\ref{fA1}, it can be concluded that the slow mode shocks are oriented very close to the radial
direction, i.e., that the angle denoted in Fig.~\ref{fA1} as $\phi$, is very small. Since the angle
$\phi$ (expressed in radians) equals to the inflow Alfv\'en Mach number reduced by the factor
$n_2/n_1$ that represents the density jump at the slow mode shock \citep[the value of $n_2/n_1$
depends on $\beta$, and should be around 2; see][]{vrssken05}, one finds that the inflow Mach
number is low, most likely in the order of 0.01 or less. [For larger values, $\phi$ would be large
enough to make the ``super-radial" widening of the ray measurable.] This may be the reason why
inflows like those shown in Fig.~\ref{f6} are rarely observed.

Note that the situation shown in Fig.~\ref{f6} is different from the previously described
``stationary'' pattern. In this case we see distinct elongated features inflowing from both sides
towards the ray axis of symmetry. Thus, these features cannot be interpreted as CS boundaries,
i.e., signatures of the slow mode shocks, but rather they represent coronal density inhomogeneities
aligned with the magnetic fieldlines. These features are inclined from the radial direction, i.e.,
show the ``super-radial" orientation. When observed in pairs, they form a V-pattern roughly
symmetric with respect to the ray axis. Thus, these substructures most likely outline the magnetic
field inflowing into the current sheet. In such a situation the angle (expressed in radians)
between the fieldline and the axis of symmetry is equal to the inflow Alfv\'en Mach number $M_A$
\citep[e.g.,][]{vrssken05}. In the event shown in Fig.~\ref{f6}, we estimate this angle to a few
degrees, so we can take $M_A\sim$\,0.02\,--\,0.05. Bearing in mind that the inflow speed was
estimated to $\sim20$~km\,s$^{-1}$, this would correspond to the Alfv\'en speed of
400\,--\,1000~km\,s$^{-1}$, which seems reasonable for this height range \citep[see,
e.g.,][]{vrs04bsplitIII}.

The main physical characteristics of post-CME rays are their mass/density excess and increased
temperature (Table I). This, as well as the morphology and flows, can be explained in terms of the
reconnection outflow-jet, being the structural element of the vertical current sheet that forms in
the wake of a CME. The measured values of the CS density and the temperature provide also an
estimate of the ambient coronal magnetic field, since the external magnetic pressure is roughly
equal to the gas pressure in the outflow region \citep[see, e.g., Appendix in][]{aurass02fms}.
Using the values $n_{CS}=5$\,--\,10\,$\times 10^7$~cm$^{-3}$ and $T_{CS}=5$\,--\,8 MK, we find for
the ambient magnetic field $B_c\sim 0.9$\,--\,1.7 gauss. Such a magnetic field is consistent with
the empirical scaling $B=0.5 (R-1)^{-1.5}$ established by \citet{D&M78}, which gives $B\sim
0.85$\,--\,1.41 gauss for the radial distance range $R=1.5$\,--\,1.7. Similarly, using the
relationship between the temperature jump at SMSs and the external plasma-to-magnetic pressure
ratio, $T_2/T_1=1+0.4/\beta$ \citep[see Eq.\,12 in][]{aurass02fms}, one can estimate the value of
$\beta$ in the inflow region. Applying $T_2/T_1\lesssim T_{CS}/T_c=2$\,--\,8, we find $\beta\gtrsim
0.4$\,--\,0.06. Such values are roughly consistent with those expected for the active region corona
in this height range \citep{gary01}.


The results of the model, based on the quasi-stationary \citet{petschek64} reconnection regime, are
consistent with the observed density excesses. At the standing slow mode shocks, formed by the CS
inflow, the plasma is heated, compressed, and accelerated/deflected to form the upward directed
reconnection jet. In this way the dense plasma from low corona is transported to larger heights,
causing the observed density excess. Model results show that the density can be increased by more
than one order of magnitude, which is consistent with observations. However, it should be noted
that the observed values depend on the assumed value of the LoS CS-depth $\lambda_0$. Our results
indicate that $\lambda_0$ is in the order of 100 Mm, and that the plane of the observed rays is
inclined at small angle with respect to the LoS.

The parameter that does not depend on the LoS CS-depth, or the CS orientation, is the number of
electrons per unit length of the CS, which we denoted in Sect.~3.3 as $N_e$. The measurements show
that $N_e(R)$ either decreases, or is approximately constant. On average, $N_e(R)$ first decreases,
to become approximately constant at $R_c\sim 4$. Given our observations, as well as the
observations by \citet{ciara08}, it seems that $R_c$ is different in different events.

Bearing in mind the equation of continuity, there are two effects that determine the radial
dependence of $N_e$. One is the radial dependence of the reconnection-outflow velocity, and another
one is the contribution of the inflowing plasma. For example, if the outflow speed would be
constant, the value of $N_e$ would be increasing monotonously with the height due to the cumulative
supply of the plasma through the SMSs (more and more inflowing plasma joining to the outflow). On
the other hand, if the inflow contribution is negligible (as it is at large heights), then a
decelerated outflow would be associated with increasing $N_e(R)$, whereas the accelerated outflow
would lead to a decreasing $N_e(R)$.

Inspecting outcomes for various model inputs, we found that regardless on the model details, in the
vicinity of the diffusion region $N_e(R)$ increases steeply due to a strong effect of the plasma
inflow. At larger heights, beyond $\sim$\,0.5\,$r_{\odot}$ above the diffusion region, the behavior
of $N_e(R)$ becomes dependent on the model input. A match with the observations (decreasing
$N_e(R)$ followed by $N_e\sim const.$) is found only for cases where the Alfv\'en velocity and the
solar wind speed increase in the given height range, and only if the diffusion region is below
$R\sim 1.4$. Given that the power-law fit presented in Fig.~\ref{f5}b quite well describes the
$N_e(R)$ dependence as $N_e\propto R^{-1}$, would imply that the outflow velocity increases
approximately as $v_{out}\propto R$. However, note that the last three data-points in
Fig.~\ref{f5}b might indicate that $N_e$ starts to increase gradually beyond $R\sim 5$. Such a
behavior is found in all of the considered model options, i.e., after the solar wind speed becomes
approximately constant and the Alfv\'en velocity starts decreasing, the value of $N_e$ gradually
increases with the radial distance.

In this respect, it should be stressed that the model results generally depend on the applied
coronal density and magnetic field model. On the other hand, the $N_e(R)$ measurements are burdened
by large errors in estimating the ray width $D$. So, since the comparison of the observed and the
calculated $N_e(R)$ might lead to ambiguous conclusions, it turns out that the most reliable
model-parameter to be compared with observations is the slope of the $n_{CS}(R)$ dependence, which
should be less steep than the slope of the ambient $n_c(R)$, regardless on the model details.
Comparing the calculated and observed slopes we find a very good correspondence.

Finally, the model predicts a decrease of the density excess in time, due to the rise of the
diffusion region height. Bearing in mind that this rise is slow, it is to be expected that the
decrease of the density excess should be slow too. Indeed, we did not detect a decrease of the ray
density in 12 hour interval analyzed in the ray of 9 January 2002. However, note that a decrease of
the density was reported by \citet{ciara08} in the event of 4 November 2003. From their Table 1 we
find that the density decreased by $\sim 40$\,--\,50\,\% in about 2 hours. According to our model,
such a decrease would correspond to an increase of the diffusion region height for $\Delta R\sim
0.1$, which would correspond to the rise speed in the order of 10~km\,s$^{-1}$. On the other hand,
given the accuracy of our density estimates ($\sim 50$\,\%), we infer that in the event of 9
January 2002 the rise speed of the diffusion region was less than 2~km\,s$^{-1}$.

\section*{Appendix}

The overall geometry of the post-CME ray, assumed to be a signature of the bifurcated reconnecting
current sheet, is shown in Fig.~\ref{f1}. The diffusion region, where the magnetic field lines
reconnect, is located at the radial distance $R_{\rm x}$. The oppositely directed field lines merge
at the velocity $v_{\rm in}(R)$, bringing into the reconnection outflow jet (the ray) the coronal
plasma of the density $n_{\rm c}(R)$. We assume that the merging velocity is faster than the local
slow-mode speed, so the slow mode shocks (SMSs) form in between the two inflows, bounding the
reconnection outflow like in the \cite{petschek64} reconnection model.

Given the spherical geometry of the corona, the flow velocity vector in the inflow region can be
represented as a superposition of the radial component (corresponding to the solar wind speed
$v_{\rm sw}$) and the ``horizontal'' component $v_{\rm in})$ (direction of the unit vector
$\hat\vartheta$ shown in the lower-right corner of Fig.~\ref{fA1}), as depicted in Fig.~\ref{f1}b.

In the stationary state the relation $\nabla\times{\bf E}=0$ must be satisfied, to provide
$\partial{\bf B}/\partial t=0$. In a spherical coordinate system (see the lower-right corner of
Fig.~\ref{fA1}) this implies $\partial(rE_{\hat\varphi})/\partial r=0$, i.e., $Rv_{\rm in}B=const.$
For the ambient coronal magnetic field $B(R)$ we take the empirical model by \cite{D&M78} and the
simple analytical model by \cite{mann99esa} [see also \cite{vrs02bsplitII}], which then defines
$v_{\rm in}(R)$.


We denote the angle between the radial direction and the slow mode shock (Fig.~\ref{fA1}) as
$\phi(R)$. Taking into account the continuity equation across the SMS
\begin{equation}
v_{\rm in}\rho_{\rm c}\Delta R = v_{\rm out}\rho_{\rm SMS}\Delta R\,\phi\,,
 \label{A1}
\end{equation}
where $v_{\rm out}\sim v_{\rm A}+v_{\rm sw}$ \citep{skender03,vrssken05}, we find
\begin{equation}
\phi=\frac{\rho_{\rm c}}{\rho_{\rm SMS}}~\frac{v_{\rm in}}{v_{\rm A}+v_{\rm sw}}\,,
 \label{A2}
\end{equation}
which defines the geometry of the SMSs.

For the coronal density $\rho_{\rm c}(R)$ we use the isothermal solar wind model by Parker
\citep[see][]{mann99} using for the temperature $T_c=1$, 1.5, and 2~MK, which provides also the
corresponding solar wind speed $v_{\rm sw}(R)$. This, together with the previously defined $B(R)$,
also defines the ambient coronal Alfv\'en velocity $v_{\rm A}(R)$ and plasma-to-magnetic pressure
ratio $\beta(R)$, both governing the jump relations at the SMS. Assuming that the guiding field is
negligible, the density and temperature jump at SMS can be expressed as:
\begin{equation}
\frac{\rho_{\rm SMS}}{\rho_{\rm c}} = \frac{5(1+\beta)}{2+5\beta}\,,
 \label{A3}
\end{equation}
\begin{equation}
\frac{T_{\rm SMS}}{T_{\rm c}} = 1+\frac{2}{5\beta}\,,
 \label{A4}
\end{equation}
\citep{aurass02fms,skender03,vrssken05}, which determines the density and the temperature of the
plasma inflowing into the reconnection outflow jet at a given $R$.

At a given radial distance $R$ the density of the reconnection outflow is determined by the flow
carried from lower heights and by the flow carried across the SMS. Bearing in mind the geometry
depicted in Fig.~\ref{fA1}, the continuity equation (mass conservation) can be written in the form:
\begin{equation}
 \rho^{\rm out}_{i+1}v^{\rm out}_{i+1}L_{i+1}d_{i+1} =
 \rho^{\rm out}_{i}v^{\rm out}_{i}L_{i}d_{i}
 +  \rho^{\rm in}_{i+1}v^{\rm in}_{i+1}\overline L_{i+1}\Delta R\,,
  \label{A5}
\end{equation}
where $\overline L_{i+1}\equiv(L_i+L_{i+1})/2$, $\rho^{\rm in}_{i+1}\equiv\rho^{\rm in}(\overline R
_{i+1})$, and $v^{\rm in}_{i+1}\equiv v^{\rm in}(\overline R _{i+1})$ are the current sheet length,
the ambient density, and the ambient velocity at $R=\overline R_{i+1}\equiv R_i+\Delta R/2$,
respectively. The term on the left-hand side represents the mass flow through the upper surface
$L_{i+1}\times d_{i+1}$, where $L$ represents the length of the current sheet perpendicular to the
plane of  Fig.~\ref{fA1}. The first term on the right-hand side represents the mass flow through
the bottom surface $L_{i}\times d_{i}$ and the second one the inflow across the SMS. The flow
speeds $v^{\rm out}$ can be estimated as $v^{\rm out}_i=v_{\rm A}(R_i)+v_{\rm sw}(R_i)$ and $v^{\rm
out}_{i+1}=v_{\rm A}(R_{i+1})+v_{\rm sw}(R_{i+1})$.

Bearing in mind the spherical geometry, the length $L$ reads
\begin{equation}
L(R)=L(R_{\rm x})\frac{R}{R_{\rm x}}\,,
 \label{A6}
\end{equation}
where $R_{\rm x}$ is the radial distance of the diffusion region and $L(R_{\rm x})$ is the current
sheet extension in the $\hat\varphi$-direction at $R_{\rm x}$, i.e., it represents the length of
the current sheet X-line. The width of the current sheet at a given radial distance can be
evaluated using
\begin{equation}
 d_{i+1}= ~d_i\,\frac{R_{i+1}}{R_i} ~+ ~\Delta R\,\phi_{i+1}\,,
  \label{A7}
\end{equation}
with $d_{\rm x} \equiv d(R_{\rm x}) \sim 0$, i.e., $d_1=\phi_1\Delta R$. The first term on the
right-hand side of Eq.~(\ref{A7}) is due to the spherical geometry (hereinafter we abbreviate
$d_i\,R_{i+1}/R_i\equiv d_{i+1}^*$, i.e., $d_{i+1}=d_{i+1}^* +\Delta R\,\phi_{i+1}$), and the
second term is due to the inclination of the SMS with respect to the local radial direction
(Fig.~\ref{fA1}).

Equations (\ref{A1}--\ref{A7}) provide evaluation of the density distribution along the post
CME-ray, $\rho_{\rm out}(R)$, for various values of diffusion region height $R_{\rm x}$. In
Fig.~\ref{fA2} we show the results obtained using the coronal magnetic field model by \cite{D&M78}
and \cite{mann99esa}, whereas for the coronal density we use the isothermal ($10^6$~K) solar wind
model by \cite{mann99}.

Let us now consider the plasma temperature along the current sheet. In the first layer above the
X-line the temperature equals the temperature $T_{\rm SMS}$ evaluated by Eq.~(\ref{A4}) at the
radial distance $R=\overline R_1\equiv R_{\rm x}+\Delta R/2$. The plasma of this temperature moves
outward and enters into the next layer through the area $d_1\times L_1$, undergoing adiabatic
cooling due to volume expansion. In the second layer it meets with the plasma which inflows into
the current sheet through the slow mode shock, where it was heated to $T_{\rm SMS}(\overline R_2)$.
The same happens successively in each further level.

Since thermal conductivity of coronal plasma is very high, the temperature can be consider as
uniform across each layer, i.e., along the magnetic field lines connecting SMSs (magnetic field
component $B_{\vartheta}$). Thus, we can write for the temperature of a given layer:
\begin{equation}
 T_{\rm CS} =\frac{m_{d}T_{d}+m_{\rm SMS}T_{\rm SMS}}{m_{d}+m_{\rm SMS}}\,.
  \label{A8}
\end{equation}
Here, $m_{d}$ and $T_{d}$ are the mass and temperature of the plasma contained in the volume
defined by the lines denoted in Fig.~\ref{fA1} as $d_i$ and $d_{i+1}^*$. Note that we have
neglected the thermal conductivity along the CS axis of symmetry, since the heat flow in this
direction is reduced due to the magnetic field component $B_{\vartheta}$, and the temperature
gradients are small due to large vertical length-scale of CS. The mass $m_{d}$ is determined by:
\begin{equation}
 m_d^{i+1} = \frac{\rho_{i+1} +\rho_i}{2} \Delta R~\frac{L_{i+1} + L_i}{2}~\frac{d_{i+1}^* + d_i}{2}\,,
  \label{A9}
\end{equation}
whereas $m_{\rm SMS}$ and $T_{\rm SMS}$ are the mass and temperature of the plasma contained in the
``triangular'' volume between the radial and the slow mode shock. The temperature $T_{\rm SMS}$ can
be evaluated by Eq.~(\ref{A4}) at the radial distance $R=\overline R_{i+1}\equiv R_i+\Delta R/2$,
whereas the mass $m_{\rm SMS}$ is determined by:
\begin{equation}
 m_{\rm SMS}^{i+1} = \rho_{\rm SMS}^{i+1}\Delta R~\frac{L_{i+1} + L_i}{2}~\frac{ \Delta R\phi_{i+1}}{2}\,.
  \label{A10}
\end{equation}
where $\rho_{\rm SMS}^{i+1}$ is evaluated by Eq.~(\ref{A3}) at the radial distance $R=\overline
R_{i+1}=R_i+\Delta R/2$.

The estimate of the temperature $T_{\rm CS}$ in a given layer still requires evaluation of the
temperature $T_d$, where we have to take into account the adiabatic expansion from the layer
``$i$'' to ``$i+1$''. The plasma contained in the volume defined by lines $d_{i-1}$ and $d_{i}$:
\begin{equation}
 V_i = \Delta R~\frac{L_{i} + L_{i-1}}{2}~\frac{d_{i} + d_{i-1}}{2}\,,
  \label{A11}
\end{equation}
expands into the volume:
\begin{equation}
 V_{i+1} = \Delta R~\frac{L_{i+1} + L_{i}}{2}~\frac{d_{i+1}^* + d_{i}}{2}
     ~+~ (L_{i+1}d_{i+1}^*v^{\rm out}_{i+1} - L_{i}d_{i}v^{\rm out}_{i})\Delta t \,.
  \label{A12}
\end{equation}
The time $\Delta t$ can be expressed as
\begin{equation}
 \Delta t =  \frac{\Delta R}{\overline v_{i+1}^{\rm out}}\,,
  \label{A13}
\end{equation}
where $\overline v_{i+1}^{\rm out} = (v_{i+1}^{\rm out}+v_{i}^{\rm out})/2$. The difference of the
first term on the right-hand-side of Eq.~({\ref{A12}) and $v_i$ represents the expansion in
$\hat\varphi$ and $\hat\vartheta$ direction, whereas the second term on the right-hand-side of
Eq.~({\ref{A12}) represents the contribution of the expansion in the $\hat r$-direction. Bearing in
mind the characteristics of the adiabatic expansion, finally we get:
\begin{equation}
 T_d^{i+1} = T_{\rm CS}^{i}~\big(\frac{V_i}{V_{i+1}}\big) ^{(\gamma-1)}\,,
  \label{A14}
\end{equation}
where $\gamma$ is the ratio of specific heats, for hydrogen plasma being equal to $\gamma=5/3$.

The system Eqs.~(\ref{A1}--\ref{A14}) can be solved starting from the first layer above the
diffusion region, and successively calculating the values in every new layer by using the values
obtained for previous layer. In this way we get the dependencies $\rho(R)$ and $T(R)$ for the
current sheet plasma.

 \begin{figure}
 \includegraphics[width=7.5cm]{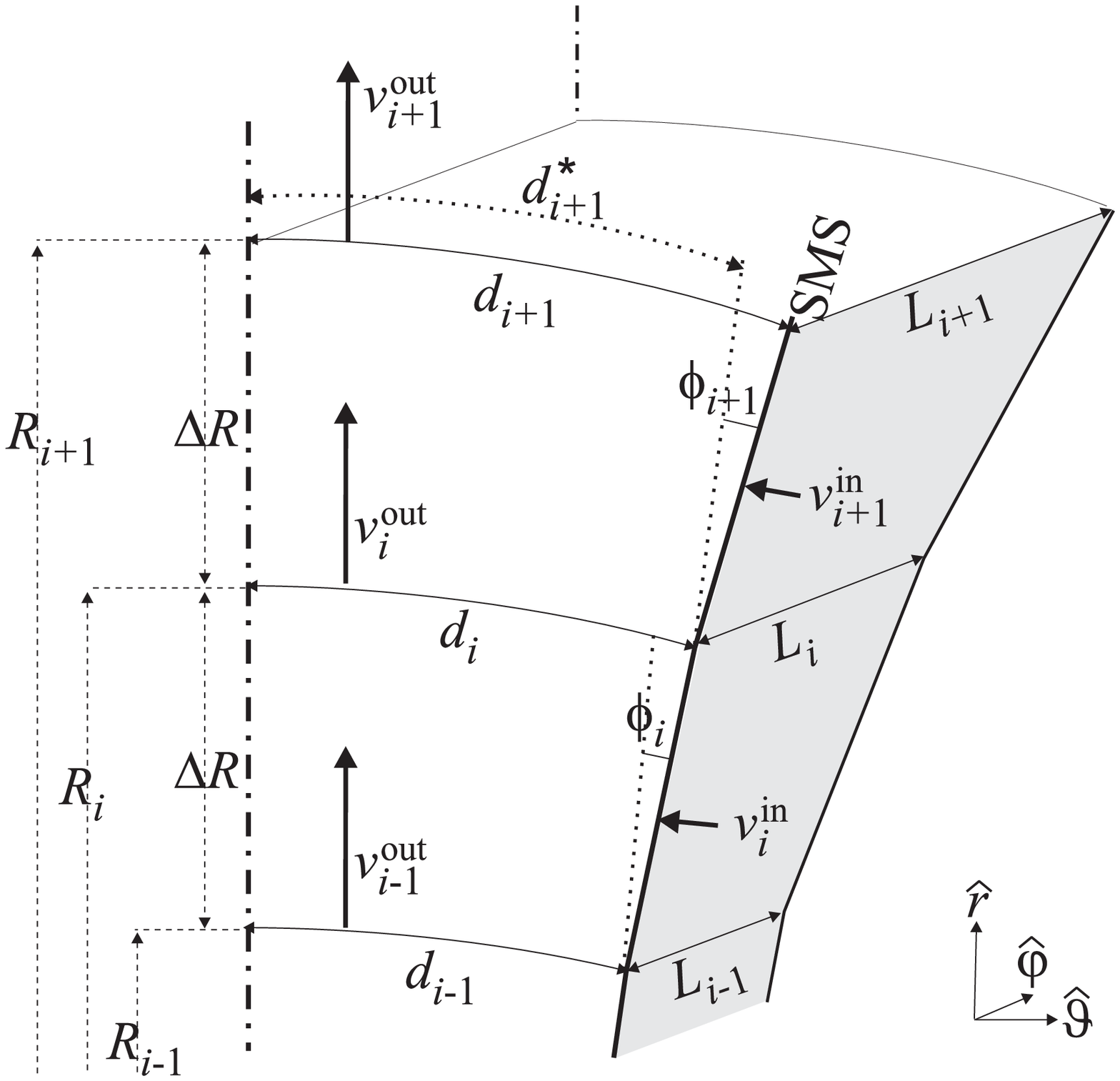}
  \centering
 \caption{
   {  Geometry of the post-CME ray model. Dash-dotted lines depict the axis of symmetry,
   shaded area represents the slow mode shock (SMS) and dotted lines the local radial direction.
   Local coordinate system is indicated in the lower-right corner. For details see the
   text in the Appendix.
 }
 \label{fA1}
 }
\end{figure}

 \begin{figure}
 \includegraphics[width=7.5cm]{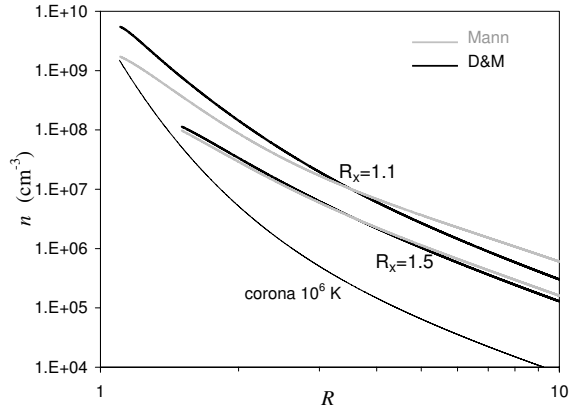}
  \centering
 \caption{
   {  Current sheet number-density calculated using the coronal magnetic field model by \cite{D&M78} and \cite{mann99esa};
   thick-black and thick-gray lines, respectively.
   The solar wind density model by \cite{mann99} is drawn by thin-gray line
   (denoted as ``corona $10^6$~K").
 }
 \label{fA2}
 }
\end{figure}

\acknowledgements
 We thank ISSI (International Space Science Institute, Bern) for the hospitality
provided to the members of the team on the Role of Current Sheets in Solar Eruptive Events where
many of the ideas presented in this work have been discussed. G.P. acknowledges support from
ASI/INAF I/015/07/0.

\bibliographystyle{aa}
\bibliography{issiref}

\end{document}